\documentclass[twocolumn,switch]{article}
\usepackage{template/arxiv}

\usepackage[utf8]{inputenc} 
\usepackage[T1]{fontenc}    
\usepackage{hyperref}       
\usepackage{url}            
\usepackage{booktabs}       
\usepackage{nicefrac}       
\usepackage{microtype}      

\usepackage{cite}
\usepackage{amsmath,amssymb,amsfonts}
\usepackage{algorithmic}
\usepackage{graphicx}
\usepackage{textcomp}

\newcommand{\Model}{CNN-QoE}
\newcommand{\ModelTCN}{TCN-QoE}

\title{Convolutional Neural Networks for Continuous QoE Prediction in Video Streaming Services}

\usepackage{authblk}

\author[1]{Tho Nguyen Duc}
\author[1]{Chanh Minh Tran}
\author[2]{Phan Xuan Tan}
\author[1]{Eiji Kamioka}
\affil[1]{Graduate School of Engineering and Science, Shibaura Institute of Technology, Japan}
\affil[2]{Deparment of Information and Communications Enginerring, Shibaura Institute of Technology, Japan}

\begin{document}

\twocolumn[ 
  \begin{@twocolumnfalse} 
  
\maketitle


\begin{abstract}
In video streaming services, predicting the continuous user's quality of experience (QoE) plays a crucial role in delivering high quality streaming contents to the user.
However, the complexity caused by the temporal dependencies in QoE data and the non-linear relationships among QoE influence factors has introduced challenges to continuous QoE prediction.
To deal with that, existing studies have utilized the Long Short-Term Memory model (LSTM) to effectively capture such complex dependencies, resulting in excellent QoE prediction accuracy.
However, the high computational complexity of LSTM, caused by the sequential processing characteristic in its architecture, raises a serious question about its performance on devices with limited computational power.
Meanwhile, Temporal Convolutional Network (TCN), a variation of convolutional neural networks, has recently been proposed for sequence modeling tasks (e.g., speech enhancement), providing a superior prediction performance over baseline methods including LSTM in terms of prediction accuracy and computational complexity.
Being inspired of that, in this paper, an improved TCN-based model, namely \Model{}, is proposed for continuously predicting the QoE, which poses characteristics of sequential data.
The proposed model leverages the advantages of TCN to overcome the computational complexity drawbacks of LSTM-based QoE models, while at the same time introducing the improvements to its architecture to improve QoE prediction accuracy.
Based on a comprehensive evaluation, we demonstrate that the proposed \Model{} model can reach the state-of-the-art performance on both personal computers and mobile devices, outperforming the existing approaches.
\end{abstract}

\keywords{
Convolutional Neural Networks \and Temporal Convolutional Network \and Quality of experience \and Video streaming}

\vspace{1cm}

  \end{@twocolumnfalse} 
] 


\section{Introduction}
\label{sec:Introduction}
For years, video streaming services have increasingly become the most dominant services on the Internet, creating an extremely huge profit for streaming service providers.
Within such a highly competitive streaming service market, service providers such as YouTube, Netflix, or Amazon must provide a sufficient video quality to satisfy the viewer's expectation, resulting in a high quality of experience (QoE).
However, video streaming services are frequently influenced by dynamic network conditions that can lead to distorted events, subsequently causing QoE deterioration.
Therefore, developing QoE models that quickly and accurately predict the user's QoE in real-time can significantly benefit QoE-aware applications.
By relying on a QoE model, for instance, a stream-switching controller designed at a client-side \cite{QABR_Chanh, QABR_DeepQ} with an adaptive bitrate selection algorithm can adaptively predict and request an optimal video quality level.
However, the continuous evaluation of QoE is challenging since it needs to capture the complex temporal dependencies in sequential QoE data and the non-linear relationships among QoE influencing factors (e.g., video quality, bitrate switching, and rebuffering) \cite{EffectSizesOfInfluenceFactors, QoEModel_TimeVaryingSubjectiveQuality, QoEModel_TVQoE_ContinuousTimeQoE, NetflixQoE}.
To deal with this challenge, a QoE prediction model which leverages Long Short-Term Memory (LSTM) \cite{QoEModel_LSTM} was introduced.
The LSTM-based QoE prediction model achieved the state-of-the-art accuracy since it is capable of capturing temporal dependencies in sequential QoE data.
However, the chain structure in the LSTM architecture requires a high computational cost for practically predicting the user's QoE due to the use of sequential processing over time.
It means that the subsequent processing steps must wait for the output from the previous ones.
This leads to an open question about the performance of the model on power-limited computers like mobile devices that may not have enough computational power to implement such QoE-aware algorithms.

Recently, Temporal Convolutional Network (TCN) \cite{Network_TCN}, a variation of Convolutional Neural Network (CNN), has emerged as a promisingly alternative solution for the sequence modeling tasks.
TCN adopts dilated causal convolutions \cite{Network_Dilated1, Network_Dilated2, Network_Dilated3} to provide a powerful way of extracting the temporal dependencies in the sequential data.
Different from LSTM, the computations in TCN can be performed in parallel, providing computational and modeling advantages.

In practical deployments, TCN convincingly outperforms canonical recurrent architectures including LSTMs and Gated Recurrent Units (GRUs) across a broad range of sequence modeling tasks \cite{Network_TCN}.
Enlightened by the great ability of TCN, in this paper, we propose an improved TCN-based model, namely QoE-CNN, for continuous QoE prediction on different viewing devices (i.e., personal computers and mobile devices).


The goal of this study is to enhance the QoE prediction accuracy while minimizing the computational complexity to support a diversity of platforms and devices.
The contributions of this paper are as follows:

\begin{itemize}
  \item
    First, \emph{\Model{}} model, an improved model of TCN for continuous QoE prediction in real-time, is proposed.
  \item
    Second, an optimal model architecture hyperparameters set for the proposed model is introduced to achieve the best QoE prediction performance.
  \item
    Third, a comprehensive evaluation of the CNN-QoE is performed across multiple QoE databases in comparison with different baseline methods.
    The results show that the CNN-QoE achieves superior performance in terms of accuracy and computational complexity on both personal computers and mobile devices.
\end{itemize}


The remainder of the paper is organized as follows:
Section \ref{sec:RelatedWork} describes the limitations of existing works for QoE modeling in video streaming.
Section \ref{sec:TCN} discusses the TCN architecture in detail.
The proposed model is presented in Section \ref{sec:ProposedModel}.
Section \ref{sec:Evaluation} and \ref{sec:Discussion} provide evaluation results of the proposed model and their discussion, respectively.
Finally, the paper is concluded in Section \ref{sec:Conclusion}.

\section{Related Work}
\label{sec:RelatedWork}
QoE modeling for video streaming services has received enormous attentions due to its critical importance in QoE-aware applications.
A number of different continuous QoE prediction models have been proposed \cite{QoEModel_TimeVaryingSubjectiveQuality, QoeModel_ShortLongTermQualityModel, QoEModel_BitrateDistribution, QoEModel_NARX_DynamicNetworks, QoEModel_AugmentedAutoregressive, QoEModel_TVQoE_ContinuousTimeQoE, QoEModel_NLSS}.
The authors in \cite{QoEModel_TimeVaryingSubjectiveQuality} modeled the time-varying subjective quality (TVSQ) using a Hammerstein-Wiener model.
The work in \cite{QoEModel_AugmentedAutoregressive} proposed a model based on the augmented Nonlinear Autoregressive Network with Exogenous Inputs (NARX) for continuous QoE prediction.
It should be noted that these models did not consider rebuffering events which usually happen in video streaming \cite{Survey_QoE, Survey_QoEModeling}.
On the other hand, the study in \cite{QoEModel_NARX_DynamicNetworks} took into account rebuffering events, perceptual video quality and memory-related features for QoE prediction.
However, the QoE prediction accuracy varied across different playout patterns.
The reason is that the model suffered from the difficulty in capturing the complex dependencies among QoE influence factors, leading to unreliable and unstable QoE prediction performances.

In order to address the above challenges, the authors in \cite{QoEModel_LSTM} proposed a QoE prediction model, namely, LSTM-QoE, which was based on Long Short-Term Memory networks (LSTM).
The authors argued that the continuous QoE is dynamic and time-varying in response to QoE influencing events such as rebuffering \cite{MeasuringQoEOfHTTP} and bitrate adaptation \cite{QoEModel_TimeVaryingSubjectiveQuality}.
To capture such dynamics, LSTM was employed and the effectiveness in modeling the complex temporal dependencies in sequential QoE data was shown.
The model was evaluated on different QoE databases and outperformed the existing models in terms of QoE prediction accuracy.
However, the computational complexity of the model was not fully inspected.
Since the recurrent structure in LSTM can only process the task sequentially, the model failed to effectively utilize the parallel computing power of modern computers, leading to a high computational cost.
Therefore, the efficiency of the model on different viewing devices with a limited computing power (i.e., mobile devices) remains an open question.

Recently, a CNN architecture for sequence modeling, Temporal Convolutional Network (TCN) \cite{Network_TCN}, was proposed.
The dilated causal convolutions \cite{Network_Dilated1, Network_Dilated2, Network_Dilated3} in TCN enable it to efficiently capture the complex dependencies in a sequential data.
The convolution operations can also be performed in parallel which effectively addresses the computational cost problem of LSTM.
Besides, TCN has been successfully employed to tackle the complex challenges in sequence modeling tasks such as speech enhancement \cite{Network_TCN}.
Therefore, in this paper, we present \Model{}, a continuous QoE prediction model based on TCN architecture, for improving the QoE prediction accuracy and optimizing the computational complexity.



\section{Temporal Convolutional Network}
\label{sec:TCN}
\begin{figure}
  \centering
  \includegraphics[width=0.95\linewidth]{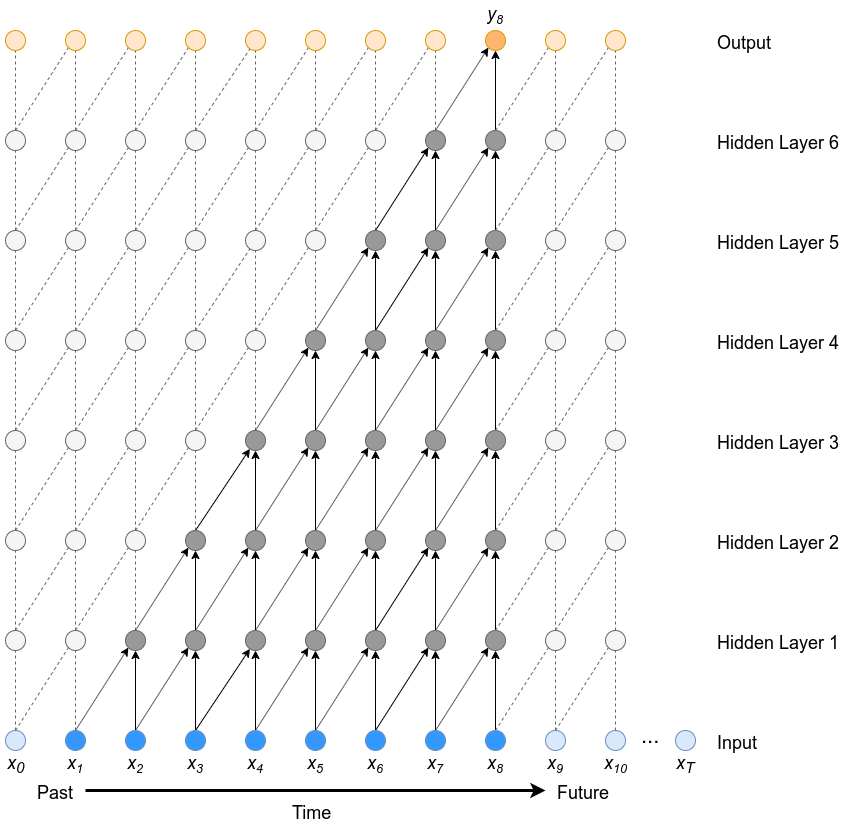}
  \caption{An illustration of a stack of causal convolution layers with the convolution filter size of $1 \times 2$.}
  \label{fig:CausalConv}
\end{figure}

\begin{figure}
  \centering
  \includegraphics[width=0.95\linewidth]{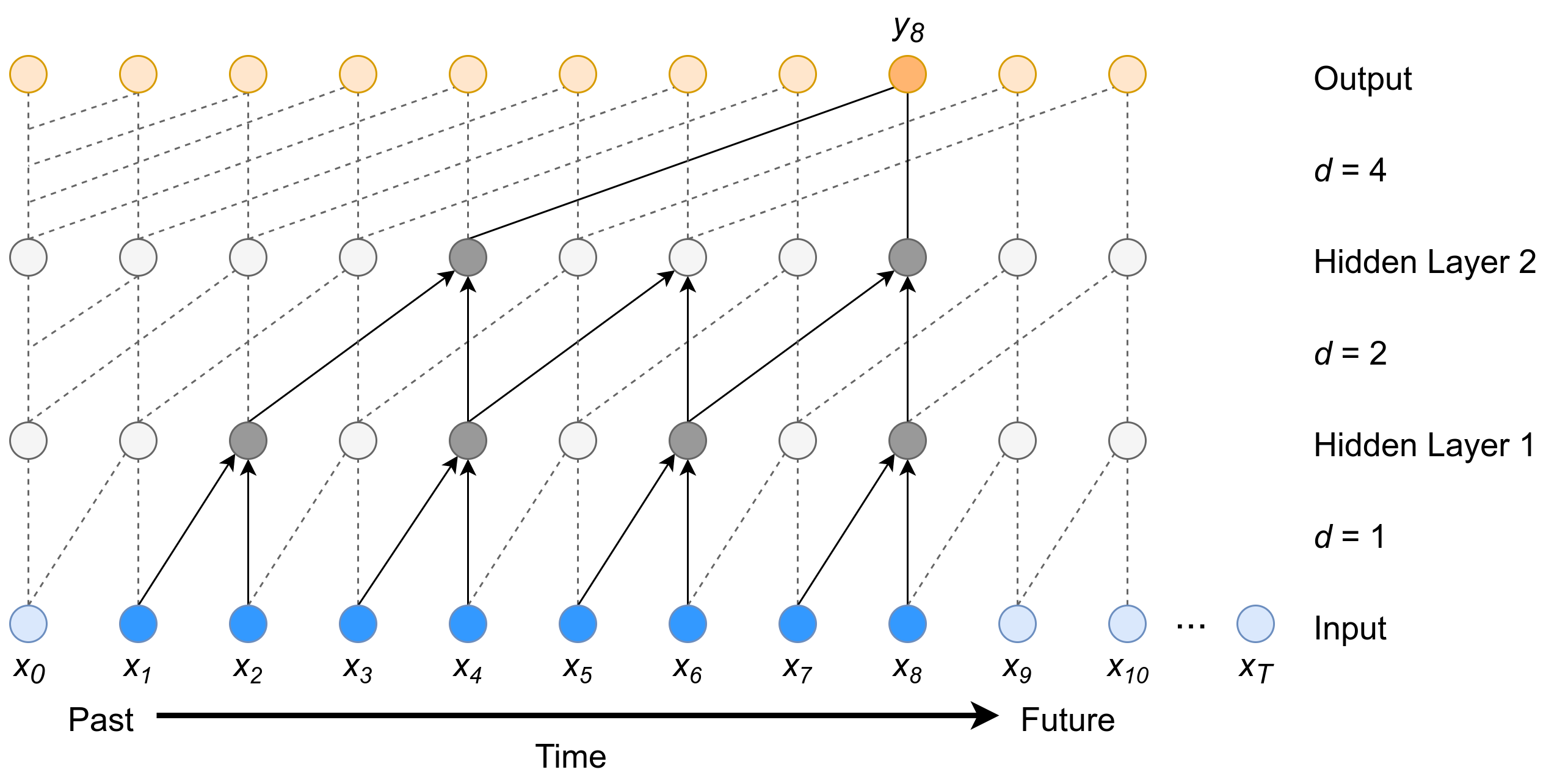}
  \caption{An illustration of a stack of dilated causal convolution layers with the convolution filter size of $1 \times 2$.}
  \label{fig:DilatedConv}
\end{figure}

\begin{figure}
  \centering
  \includegraphics[width=0.9\linewidth]{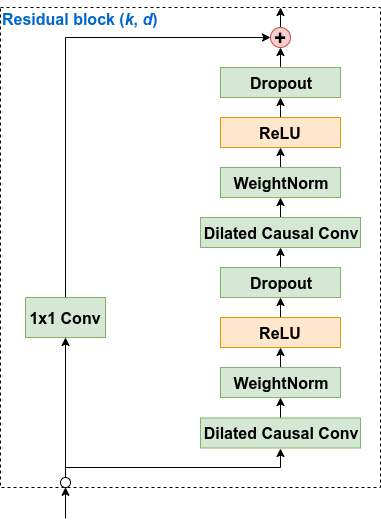}
  \caption{The residual block in TCN architecture.}
  \label{fig:TCN_residualBlock}
\end{figure}

In this section, TCN architecture is briefly discussed to summarize its advantages and disadvantages in sequence modeling tasks. Thereby, the conclusions of this section will be the crucial foundation for the subsequent improvements proposed in CNN-QoE, which are stated in section \ref{sec:ProposedModel}.


\subsubsection{1D Convolutions}

CNN was traditionally designed to operate on two dimensions (2D) data such as images.
An input image is passed through a series of 2D convolution layers.
Each 2D convolution applies and slides a number of 2D filters through the image.
To adapt CNN for time-series data, TCN utilizes 1D convolution where the filters exhibit only one dimension (time) instead of two dimensions (width and height).
Concretely, a time-series input is convolved with a filter size of $1 \times k$.

Furthermore, 1D convolutions are well-suited for real-time tasks due to their low computational requirements.
1D convolutions require simple array operations rather than matrix operations, hence, the computational complexity is significantly reduced in comparison with 2D convolutions.
In addition, the convolution operations allow fully parallel processing, resulting in a significant improvement of computational speed.


\subsubsection{Causal Convolutions}

A \textit{causal convolution} is a convolution layer to ensure there is no information "leakage" from future into past.
In other words, given an time-series input $x_0,..., x_{T}$, the predicted output $\widehat{y}_t$ at a time instant $t$ depends only on the inputs at time $t$ and earlier $\mathbf{x}_t, \mathbf{x}_{t-1}, ...,\mathbf{x}_{t-r+1}$.
For instance, as illustrated in Fig. \ref{fig:CausalConv}, the predicted $\widehat{y}_8$ is computed by a combination of the inputs $x_1, ..., x_8$.
It can be observed that, in order to achieve a long effective history size or a large receptive field size, an extremely deep network or very large filters are needed, which significantly increases the model computational complexity.
Thus, TCN architecture utilizes dilated causal convolutions rather than causal convolutions.
The advantages and disadvantages of dilated causal convolutions are discussed below.


\subsubsection{Dilated Causal Convolutions}

TCN adopts a dilated causal convolution comprising of the causal and the dilated convolutions.
The causal convolution has already been described in the previous subsection.
Meanwhile, dilated convolution \cite{Network_Dilated1, Network_Dilated2, Network_Dilated3} is a convolution where the convolution filter is applied to a larger area than its length by skipping input values with several steps.
Therefore, the dilated causal convolution can effectively allow the network to operate on a larger scale than the one with a normal convolution while ensuring that there is no leakage of information from the future to the past.
The dilated causal convolution is defined as:

\begin{equation}
  D(t) = \sum_{i=0}^{k-1} f(i) \cdot \mathbf{x}_{t-d \cdot i}
  \label{eqn:TCN_dilatedConv}
\end{equation}
where, $d$ is the dilation factor, $f$ is a filter size of $1 \times k$.
$d$ exponentially increases with the depth of the network (i.e., $d = 2^l$ at layer $l$ of the network).
For instance, given the network with $L$ layers of dilated causal convolutions $l = 1, ..., L$, the dilation factors exponentially increase by a factor of $2$ for every layer:

\begin{equation}
  d \in [2^0, 2^1, ..., 2^{L-1}]
  \label{eqn:TCN_dilatedFactor}
\end{equation}

Fig. \ref{fig:DilatedConv} depicts a network with three dilated causal convolutions for dilations 1, 2, and 4.
Using the dilated causal convolutions, the model is able to efficiently learn the connections between far-away time-steps in the time series data.
Moreover, as opposed to causal convolutions in Fig. \ref{fig:CausalConv}, the dilated causal convolutions require fewer layers even though the receptive field size is the same.
A stack of dilated causal convolutions enables the network to have a very large receptive field with just a few layers, while preserving the computational efficiency.
Therefore, dilated causal convolutions reduce the total number of learnable parameters, resulting in more efficient training and light-weight model.

However, the dilated causal convolutions have problem with local feature extraction.
As shown in Fig. \ref{fig:DilatedConv}, it can be seen that the filter applied to the time-series input is not overlapped due to the skipping steps of the dilation factor.
As long as the dilation factor increases, the feature is extracted from only far-apart time-steps, but not from adjacent time-steps.
Therefore, the local connection among adjacent time-steps is not fully extracted at higher layers.


\subsubsection{Residual Block}

The depth of the model is important for learning robust representations, but also comes with a challenge of vanishing gradients.
The residual block has been found to be an effective way to address this issue and build very deep networks \cite{Network_Residual1}.
A residual block contains a series of transformation functions $\emph{F}$, whose outputs are added to the input $\mathbf{x}$ of the block:

\begin{equation}
  o = Activation(x+\emph{F}(x))
\end{equation}

The residual block is used between each layer in TCN to speed up convergence and enable the training of much deeper models.
The residual block for TCN is shown in Fig. \ref{fig:TCN_residualBlock}.
It consists of dilated causal convolution, ReLU activation function \cite{Network_Relu}, weight normalization \cite{Network_WeightNorm1}, and spatial dropout \cite{Network_Dropout1} for regularization.
Having two layers of dilated causal convolution in the TCN's residual block is suitable for complex challenges such as speech enhancement \cite{Network_TCN}.
Compared with speech signal data, sequential QoE data is much simpler.
That is to say, the two layers of dilated causal convolution are redundant and are not optimal for the QoE prediction problem.



In TCN architecture, equations (\ref{eqn:TCN_dilatedConv}) and (\ref{eqn:TCN_dilatedFactor}) suggest that the TCN model heavily depends on the network depth $L$ and the filter size $k$.

\section{Proposed QoE Prediction Model}
\label{sec:ProposedModel}
In this section, the proposed model \Model{} is introduced to leverage the advantages and handles the problems of the TCN architecture \cite{Network_TCN} in QoE prediction tasks for video streaming services.
The main objective of our study is to enhance the QoE prediction accuracy and minimize the computational complexity.

Let $\mathbf{x}_t$ be a vector of input features at a time instant $t$ within a streaming session of $T$ seconds.

Let $y_t$ and $\widehat{y}_t$ be the subjective and the predicted QoE at a time instant $t$, respectively.
In order to predict the subjective QoE continuously at any given time instant $t$, the following nonlinear function must be considered:

\begin{equation}
  \widehat{y}_t = g(\mathbf{x}_t, \mathbf{x}_{t-1}, ..., \mathbf{x}_{t-r+1}), \quad 0 \leq t \leq T
\end{equation}

where $r$ is the number of lags in the input.
To learn the nonlinear function $g(\cdot)$, the \Model{} model is presented.


In the following subsections, the proposed architecture employed for the \Model{} model is discussed in detail.
The model architecture hyperparameters are then analyzed to find the optimal values which can improve the QoE prediction accuracy, while minimizing the computational complexity.


\subsection{Proposed Model Architecture}
\label{sec:ProposedModel_Architecture}
\begin{figure}
  \centering
  \includegraphics[width=0.95\linewidth]{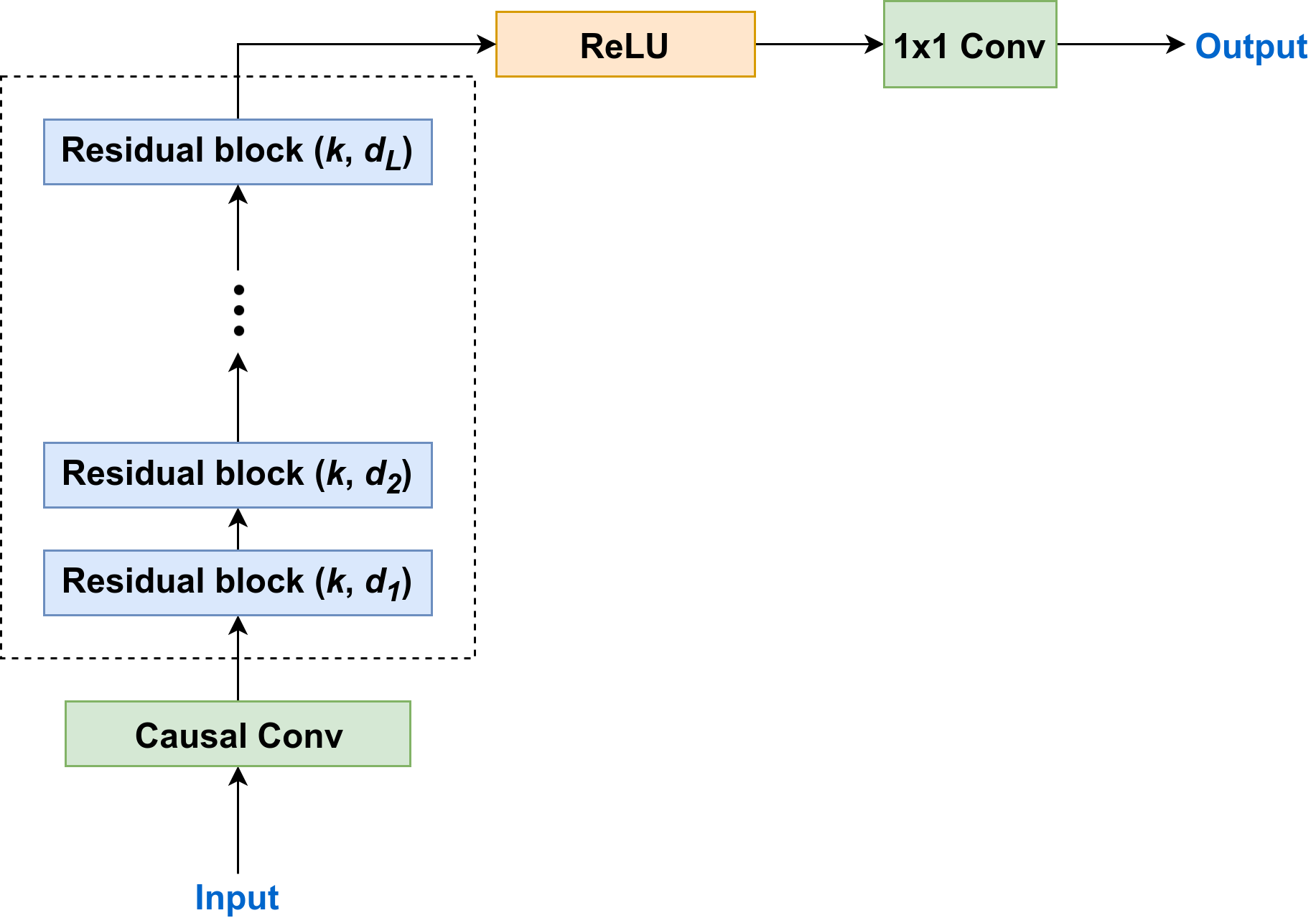}
  \caption{The proposed \Model{} architecture.}
  \label{fig:ProposedArchitecture}
\end{figure}

\begin{figure}
  \centering
  \includegraphics[width=0.95\linewidth]{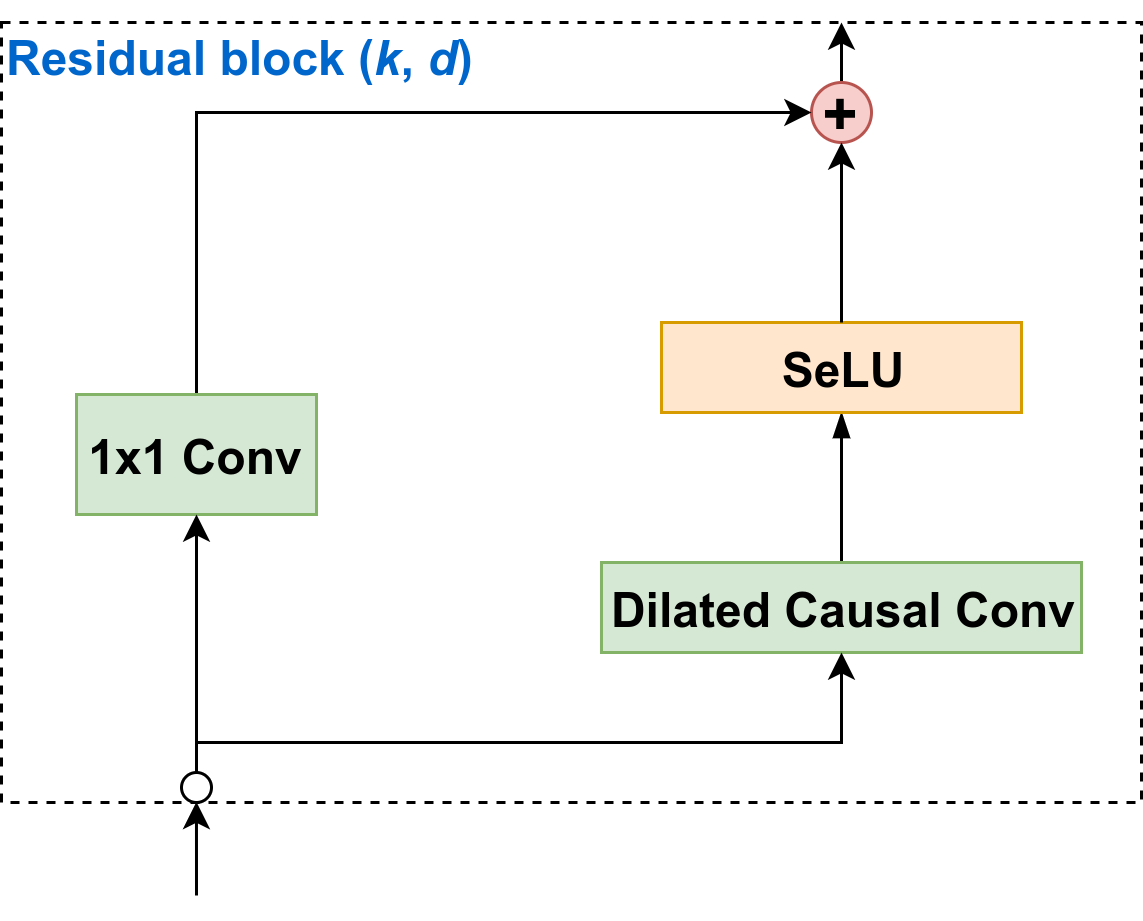}
  \caption{The proposed residual block used in the proposed architecture.}
  \label{fig:ProposedResidualBlock}
\end{figure}

Fig. \ref{fig:ProposedArchitecture} illustrates the overview of the \Model{}'s architecture.
The \Model{} leverages the advantages of 1D convolutions, dilated causal convolutions and residual block in TCN architecture.
To adapt the TCN to QoE prediction tasks, a number of improvements must be done, which are:

\begin{itemize}
  \item An initial causal convolution layer is added to the input and then connects to residual block which includes a dilated causal convolution layer.
  \item The residual block is simplified by leveraging the advantages of Scaled Exponential Linear Units (SeLU) activation function \cite{SeLU}.
\end{itemize}
These distinguishing characteristics are discussed as below.


\subsubsection{Causal convolution to extract local features}

The architecture of the proposed model comprises of one causal convolution layer and a stack of dilated causal convolutions, while the TCN consists of only a number of dilated causal convolutions.
A causal convolution layer is added between the input time-series and the first residual block as shown in Fig. \ref{fig:ProposedArchitecture}.
This causal convolution layer can extract the local features of the adjacent time-steps in the sequential QoE data.
Afterward, the following dilated causal convolution layers are leveraged to extract the global features between far-apart time steps.
These layers help the model to learn the most informative features in the time series input, resulting in higher accuracy.


\subsubsection{SeLU activation function}

Activation function plays an important role in allowing the model to learn non-linear representations of the input features.
When training a deep learning model, the vanishing and exploding gradient are the most challenging problems that prevent the network from learning the optimal function $g(\cdot)$.
The TCN model gets rid of these problems by integrating
ReLU activation function \cite{Network_Relu},
weight normalization \cite{Network_WeightNorm1} and dropout \cite{Network_Dropout1} layer as shown in Fig. \ref{fig:TCN_residualBlock}.
In the proposed \Model{}, those layers are replaced with the SeLU to leverage its advantages and simplify the residual block as shown in Fig. \ref{fig:ProposedResidualBlock}.
SeLU is a self-normalizing neural network.
It converges to zero mean and unit variance when propagated through multiple layers during network training, thereby making it unaffected by vanishing and exploding gradient problems.
Moreover, SeLU also solves the "dying ReLU" problem where the ReLU function always outputs the same value of 0 for any input, so the gradient descent is not able to alter the learnable parameters.
At the same time, SeLU also reduces the training time and learns robust features more efficiently than other networks with normalization techniques, such as weight normalization \cite{SeLU}.
SeLU activation function described as follow \cite{SeLU}:

\begin{equation}
\label{eq:SeLU}
  SeLU(x) = \lambda \begin{pmatrix}
    x, & \textit{if } x > 0 \\ 
    \alpha \textit{exp}(x) - \alpha, & \textit{if } x \leq 0 
  \end{pmatrix}
\end{equation}


where $\alpha = 1.67733$ and $\lambda = 1.0507$. These are the same values as the ones proposed in \cite{SeLU}.


\subsection{Architecture Hyperparameters Selection}
\label{sec:ProposedModel_Parameters}

\begin{table*}[t]
  \centering
  
  \begin{tabular}{|c|l|c|}
  \hline
  Architecture \par Hyperparameters & Description & Derived Value \\ \hline
  $r$ & Receptive field size & 8\\ \hline
  $k$ & Filter size & 2\\ \hline
  $L$ & Number of dilated causal convolution layers & 3\\ \hline
  $n$ & Number of filters on each convolution layer & 32\\ \hline
\end{tabular}

  \caption{Hyperparameters for the best performance model.}
  \label{tbl:ModelHyperparameters}
\end{table*}

When training the model, an adequate set of architecture hyperparameters must be selected to achieve the best performance.
The proposed model consists of $L$ residual block layers, each layer contains a dilated causal convolution with a filter size of $1 \times k$, as shown in Fig. \ref{fig:ProposedArchitecture} and \ref{fig:ProposedResidualBlock}.
Each dilated convolution layer has a dilation factor $d$ doubled at each layer up, as shown in (\ref{eqn:TCN_dilatedFactor}).
The proposed model depends on the network depth $L$ and the filter size $k$.
These hyperparameters control the trade-off between QoE prediction accuracy and computational complexity of the model.
To effectively optimize the hyperparameters, it is important to set a boundary for the space of possible hyperparameter values.

The user's QoE is mostly affected by the recent experiences, also known as the recency effect \cite{Recency, NetflixQoE, LFOVIA}.
The recency effect gradually decreases within 15 to 20 seconds \cite{NetflixQoE, LFOVIA} after distorted events (e.g., bitrate fluctuations, rebuffering events).
Therefore, the effective history or the receptive field size $r$ of the model cannot be larger than 20 time-steps

\begin{equation}
  r \leq 20
  \label{eqn:TCN_receptiveField_limit}
\end{equation}

Moreover, the receptive field depends on the number of dilated causal convolution layers $L$ and the filter size $k$.
For example, with $l \in [1, L]$, the receptive field $r$ can be determined by (\ref{eqn:TCN_receptiveField_2}) \cite{Network_Wavenet, Network_TCN}

\begin{equation}
 r = 2^{L} \mathrm{, if} \quad k = 2
 \label{eqn:TCN_receptiveField_2}
\end{equation}

or (\ref{eqn:TCN_receptiveField_3}) \cite{Network_ReceptiveField_3}

\begin{equation}
  r = 2^{L + 1} - 1 \mathrm{, if} \quad k = 3
  \label{eqn:TCN_receptiveField_3}
\end{equation}

Fig. \ref{fig:DilatedConv} shows an example of a three-layer ($L=3$) dilated convolutional network.
In this figure, given the filter size of $1 \times 2$ ($k = 2$), the receptive field is computed by $r = 2^{3} = 8$.
From (\ref{eqn:TCN_receptiveField_limit}), (\ref{eqn:TCN_receptiveField_2}), and (\ref{eqn:TCN_receptiveField_3}), the range of $L$ values can easily be defined $L \in [2, 3, 4]$.

In a 1D convolution, the number of filters $n$ is also important to effectively extract the information from the inputs.
To minimize the computation complexity of the model, the range of $n$ is set to $n \in \{16, 32, 64\}$.
We conduct a simple grid-search of the model architecture hyperparameters with $k \in [2, 3]$, $L \in [2, 3, 4]$, and $n \in \{16, 32, 64\}$.
Table \ref{tbl:ModelHyperparameters} shows the values of $r$, $k$, $L$, and $n$ that achieves the best performance.


\section{Performance Evaluation}
\label{sec:Evaluation}

\begin{table*}[!t]
  \centering
  \begin{tabular}{|l|c|c|c|c|c|}
\hline
\multicolumn{1}{|c|}{Database}    & Device Type & Rebuffering Events & Bitrate Fluctuations & Duration & QoE Range \\ \hline
LFOVIA Video QoE Database                  & TV                   & yes                  & yes                       & 120 seconds       & {[}0, 100{]}       \\ \hline
LIVE Mobile Stall Video Database II        & Mobile               & yes                  & no                        & 29-134 secs       & {[}0, 100{]}       \\ \hline
LIVE Netflix Video QoE Database            & Mobile               & yes                  & yes                       & at least 1 minute & {[}-2.28, 1.53{]}  \\ \hline
\end{tabular}

  \caption{An overview of the four public QoE databases used in the proposed model evaluation.}
  \label{tbl:Database}
\end{table*}

In this section, we evaluate the performance of the \Model{} in terms of QoE prediction accuracy and computational complexity.
The evaluation is performed by comparing the proposed model with numerous baseline models across multiple databases on both personal computers and mobile devices.
In the following subsections, firstly, the employed four input features for QoE prediction are described, followed by a brief explanation of baseline models.
Then, the evaluation results on accuracy and computational complexity are presented.
Finally, the overall performance of the proposed model is discussed to illustrate its capability for real-time QoE prediction.


\subsection{Input Features for QoE Prediction}
\label{sec:Evaluation_InputFeatures}
  
  


Video streaming users are sensitively affected by the video quality, known as \textit{short time subjective quality} (STSQ) \cite{QoEModel_TimeVaryingSubjectiveQuality}.
STSQ is defined as the visual quality of video being rendered to the user and can be predicted using any of the robust video quality assessment (VQA) metrics, such as Spatio-Temporal Reduced Reference Entropic Differences (STRRED) \cite{STRRED}, Multi-Scale Structural Similarity (MS-SSIM) \cite{MSSSIM}, Peak Signal to Noise Ratio (PSNR) \cite{PSNR}, etc.
Recent experiments have demonstrated that STRRED is a robust and high-performing VQA model when being tested on a very wide spectrum of video quality datasets, on multiple resolution and device types \cite{FeaturePredictionQoE, QoEModel_NLSS, QoEModel_LSTM}.
Therefore, in this paper, STRRED is utilized to measure the STSQ.

Rebuffering greatly impacts the user's QoE \cite{StallingEvents}.
Therefore, rebuffering information such as rebuffering length, rebuffering position and the number of rebuffering events must be investigated.
As a result, two rebuffering-related inputs are employed in this paper.
Firstly, \textit{playback indicator} (PI) \cite{QoEModel_NARX_DynamicNetworks, QoEModel_NLSS, QoEModel_LSTM} is defined as a binary continuous-time variable, specifying the current playback status, i.e., $1$ for rebuffering and $0$ for normal playback.
Secondly, as the user's annoyance increases whenever a rebuffering event occurs \cite{StallingEvents}, the \textit{number of rebuffering events} (NR) happened from the start to the current time instant of the session is considered.

Besides, the user's QoE is also affected by memory factors.
For example, more recent experiences have larger impacts on the user's perceived video quality, known as the recency effect \cite{Recency, NetflixQoE, LFOVIA}.
To capture the relation between the recency effect and the user's QoE, \textit{time elapsed since the last video impairment} (i.e., bitrate switch or rebuffering occurrence) \cite{QoEModel_NARX_DynamicNetworks, QoEModel_NLSS, QoEModel_LSTM}, denoted as TR, is utilized.


\subsection{Baseline Models}
\label{sec:Evaluation_BaselineModels}
To evaluate the QoE prediction accuracy of the proposed model on personal computers, the comparison with the state-of-the-art QoE models comprising of LSTM-QoE \cite{QoEModel_LSTM}, NLSS-QoE \cite{QoEModel_NLSS}, SVR-QoE \cite{LFOVIA}, and NARX \cite{QoEModel_NARX_DynamicNetworks} will be performed.
It is worth noting that we also make a comparison with the original TCN model, or \ModelTCN{} for short, in the QoE prediction task.
The \ModelTCN{} model uses the same network hyperparameters and input features with ones described in Section \ref{sec:ProposedModel_Parameters} and \ref{sec:Evaluation_InputFeatures}.

To evaluate the QoE prediction accuracy and computational complexity of the proposed model on mobile devices, we focus on the comparison with deep learning-based QoE prediction models since they achieve exceptionally higher accuracy.
Particularly, LSTM-QoE \cite{QoEModel_LSTM} and \ModelTCN{} are utilized in the comparison.
It is important to note that the LSTM-QoE \cite{QoEModel_LSTM} model hyperparameters are employed as reported in its respective works in order to ensure a fair comparison.



\subsection{Accuracy}
\label{sec:EvaluationAccuracy}

  \subsubsection{Databases}
  \label{sec:EvaluationAccuracy_Databases}
  There are three public QoE databases used for the evaluation of QoE prediction accuracy, including LFOVIA Video QoE Database \cite{LFOVIA}, LIVE Netflix Video QoE Database \cite{NetflixQoE}, and LIVE Mobile Stall Video Database II \cite{StallingEvents}.
The descriptions of these databases are summarized in Table \ref{tbl:Database}.

To evaluate the QoE prediction accuracy on personal computers, the evaluation procedures performed on each database are described as follows:
\begin{itemize}
  \item
    LFOVIA Video QoE Database \cite{LFOVIA} consists of 36 distorted video sequences of 120 seconds duration.
    The training and testing procedures are performed on this database in the same way as the one described in \cite{QoEModel_LSTM}.
    The databases are divided into different train-test sets.
    In each train-test sets, there is only one video in the testing set, whereas the training set includes the videos that do not have the same content and playout pattern as the test video. 
    Thus, there are 36 train-test sets, and 25 of 36 videos are chosen for training the model for each test video.

  \item
    LIVE Netflix Video QoE Database \cite{NetflixQoE}:
    The same evaluation procedure as described for LFOVIA Video QoE Database is employed.
    There are 112 train-test sets corresponding to each of the videos in this database.
    In each train-test set, the training set consists of 91 videos out of a total of 112 videos in the database (excludes 14 with the same playout pattern and 7 with the same content).
  
  \item
    LIVE Mobile Stall Video Database II \cite{StallingEvents}:
    The evaluation procedure is slightly different from the one applied to the above databases.
    Firstly, 174 test sets corresponding to each of 174 videos in the database are created.
    For each test set, since the distortion patterns are randomly distributed across the videos, randomly 80\% videos from the remaining 173 videos are then chosen for training the model and perform evaluation over the test video.

\end{itemize}

To evaluate the QoE prediction accuracy on mobile devices, for simplicity, only the LFOVIA Video QoE Database \cite{LFOVIA} is utilized to train and test the proposed model.
In this experiment, the set of 36 distorted videos in the database are divided into training and testing sets with a training:test ratio of 80:20.
Thus, there are 28 videos in the training set and 8 videos in the testing set.

  \subsubsection{Evaluation Settings}
  \label{sec:EvaluationAccuracy_Settings}

To evaluate the accuracy of the proposed model, the model hyperparameter sets and input features are used as described in Section \ref{sec:ProposedModel_Parameters} and \ref{sec:Evaluation_InputFeatures}, respectively.
The QoE prediction performance of the proposed model is first compared with baseline models described in Section \ref{sec:Evaluation_BaselineModels} on personal computers.
Then, on mobile devices, the comparison with other deep learning-based QoE prediction models (\ModelTCN{} and LSTM-QoE \cite{QoEModel_LSTM}) is focused on. To provide a fair comparison, the \ModelTCN{} and LSTM-QoE \cite{QoEModel_LSTM} models are also trained and tested on LFOVIA Video QoE Database with the training:test ratio of 80:20.
The implementations of \Model{}, \ModelTCN{}, and LSTM-QoE \cite{QoEModel_LSTM} are based on Keras library \cite{Lib_Keras} with the Tensorflow  \cite{Lib_Tensorflow} backend.

\begin{figure}
  \centering
  \includegraphics[width=0.95\linewidth]{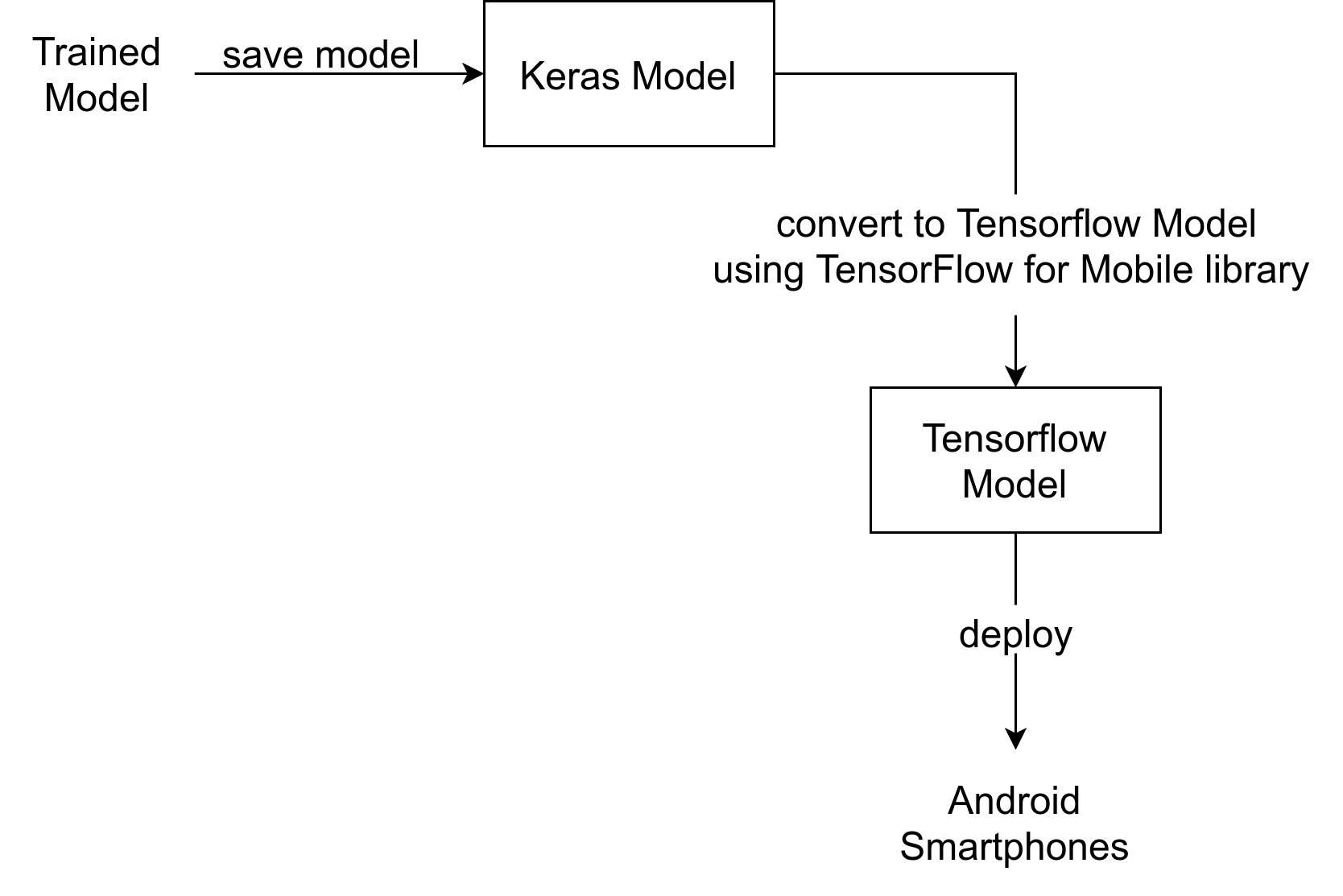}
  \caption{A diagram illustrating the steps to convert the \Model{} into a mobile device model for Android smartphones. The trained model is saved as a Keras model. The TensorFlow model is then extracted from the Keras model using "TensorFlow for Mobile" library and deployed on Android smartphones.}
  \label{fig:MobileFlow}
\end{figure}

In order to access the accuracy of the deep learning-based QoE prediction models on mobile devices (i.e., \Model{}, \ModelTCN{}, and LSTM-QoE \cite{QoEModel_LSTM}), Android smartphones are utilized for evaluation since the Android is the most popular mobile operating system \footnote{https://gs.statcounter.com/os-market-share/mobile/worldwide} all over the world.
To do so, after training phase, the trained models must be converted to the models that can be executed on the Android environment (as shown in Fig. \ref{fig:MobileFlow}).

  \subsubsection{Evaluation Criteria}
  \label{sec:EvaluationAccuracy_Criteria}
  In this paper, three evaluation metrics, namely, Pearson Correlation Coefficient (PCC), Spearman Rank Order Correlation Coefficient (SROCC) and Root Mean Squared Error (RMSE) are considered for QoE prediction accuracy assessment.
The SROCC measures the monotonic relationship, while PCC measures the degree of linearity
between the subjective and the predicted QoE
For PCC and SROCC, a higher value illustrates a better result, while for the RMSE, the lower value is better.

  \subsubsection{Results}
  \label{sec:EvaluationAccuracy_Results}
  \begin{figure*}[t]
  \centering
  \includegraphics[width=0.95\linewidth]{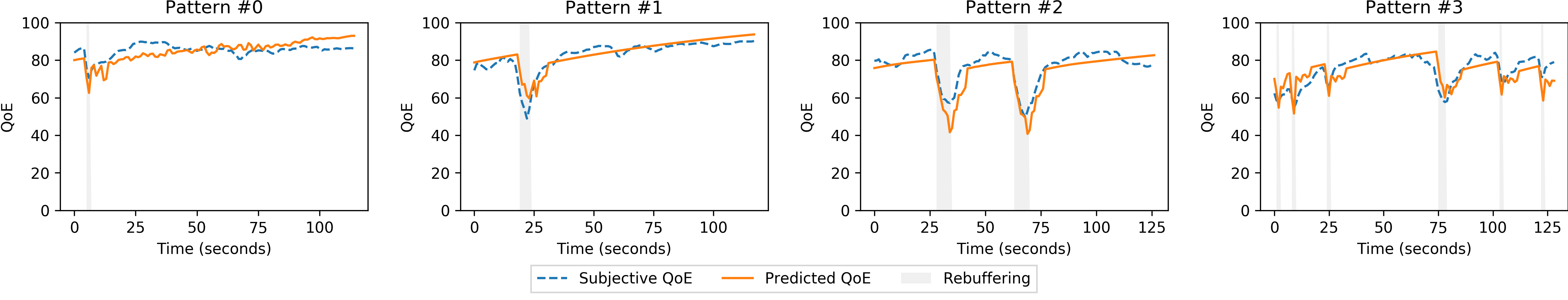}
  \caption{QoE prediction performance of the \Model{} over the LFOVIA QoE Database.}
  \label{fig:ResultLfovia}
\end{figure*}

\begin{figure*}[t]
  \centering
  \includegraphics[width=0.95\linewidth]{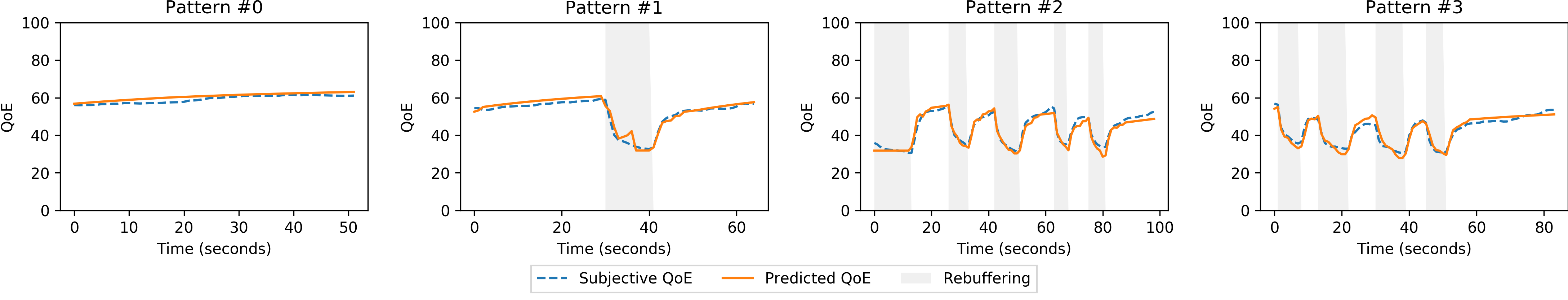}
  \caption{QoE prediction performance of the \Model{} over the LIVE Mobile Stall II Video Database.}
  \label{fig:ResultLive}
\end{figure*}

\begin{figure*}[t]
  \centering
  \includegraphics[width=0.95\linewidth]{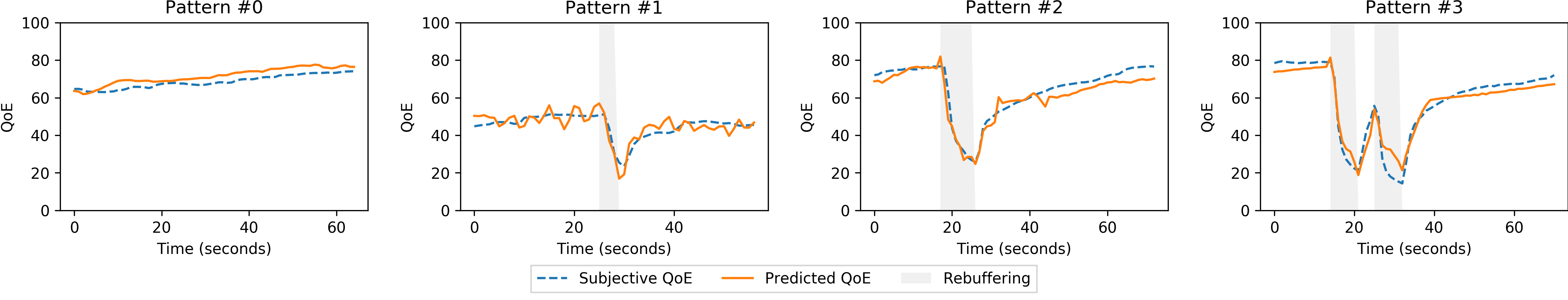}
  \caption{QoE prediction performance of the \Model{} over the LIVE Netflix Video QoE Database.\label{fig:ResultNetflix}}
\end{figure*}

\begin{table}[htb]
  \centering

  \begin{tabular}{|l|c|c|c|}
    \hline
    & PCC & SROCC & RMSE\\ \hline
    \Model{} & \textbf{0.820} & \textbf{0.759} & \textbf{4.81} \\ \hline
    \ModelTCN{}       & 0.670 & 0.732 & 5.47 \\ \hline
    LSTM-QoE \cite{QoEModel_LSTM}        & 0.800 & 0.730 & 9.56 \\ \hline
    NLSS-QoE \cite{QoEModel_NLSS}        & 0.767 & 0.685 & 7.59 \\ \hline
    SVR-QoE \cite{LFOVIA}         & 0.686 & 0.648 & 10.44 \\ \hline
  \end{tabular}

  \vspace{0.35cm}
  \caption{QoE prediction performance of the \Model{} over the LFOVIA Video QoE Database.}
  \label{tbl:Accuracy_LFOVIA}
\end{table}

\begin{table}[htb]
  \centering

  \begin{tabular}{|l|c|c|c|}
    \hline
    & PCC & SROCC & RMSE\\ \hline
    \Model & \textbf{0.892} & \textbf{0.885} & \textbf{5.36} \\ \hline
    \ModelTCN       & 0.667 & 0.603 & 9.71 \\ \hline
    LSTM-QoE \cite{QoEModel_LSTM}        & 0.878 & 0.862 & 7.08 \\ \hline
    NLSS-QoE    \cite{QoEModel_NLSS}    & 0.680 & 0.590 & 9.52 \\ \hline
  \end{tabular}
  \vspace{0.35cm}
  \caption{QoE prediction performance of the \Model{} over the LIVE Mobile Stall Video Database II. Boldface indicates the best result.}
  \label{tbl:Accuracy_LiveMobileStall}
\end{table}

\begin{table}[htb]
  \centering

  \begin{tabular}{|l|c|c|c|}
    \hline
    & PCC & SROCC & RMSE\\ \hline
    \Model & \textbf{0.848} & \textbf{0.733} & \textbf{6.97} \\ \hline
    \ModelTCN       & 0.753 & 0.720 & 7.62 \\ \hline
    LSTM-QoE \cite{QoEModel_LSTM}       & 0.802 & 0.714 & 7.78 \\ \hline
    NLSS-QoE \cite{QoEModel_NLSS}        & 0.655 & 0.483 & 16.09 \\ \hline
    NARX \cite{QoEModel_NARX_DynamicNetworks}           & 0.621 & 0.557 & 8.52\\ \hline
  \end{tabular}
  \vspace{0.35cm}
  \caption{QoE prediction performance of the \Model{} over the LIVE Netflix Video QoE Database. Boldface indicates the best result.}
  \label{tbl:Accuracy_LiveNetflix}
\end{table}

\begin{table*}[htb]
  \centering
    \begin{tabular}{|l|c|c|c|c|c|c|}
    \hline
    & \multicolumn{2}{l|}{PCC} & \multicolumn{2}{l|}{SROCC} & \multicolumn{2}{l|}{RMSE} \\ \hline
    & PC & Mobile & PC & Mobile & PC & Mobile \\ \hline
    \Model{}    & \textbf{0.873} & \textbf{0.873} & \textbf{0.878} & \textbf{0.878} & \textbf{5.27} & \textbf{5.27} \\ \hline
    \ModelTCN{} & 0.731 & 0.731 & 0.792 & 0.792 & 6.71 & 6.71 \\ \hline
    LSTM-QoE \cite{QoEModel_LSTM} & 0.863 & 0.259 & 0.822 & 0.650 & 6.82 & 22.57 \\ \hline
  \end{tabular}

  \caption{A comparison of the \Model{}'s QoE prediction performance on personal computer and mobile device over LFOVIA Video QoE Database with 80/20 split.}
  \label{tbl:Accuracy_ComparePCvsMobile}
\end{table*}

Figs. \ref{fig:ResultLfovia}, \ref{fig:ResultLive} and \ref{fig:ResultNetflix} illustrate the QoE prediction performance over the three databases using the proposed \Model{} model on personal computers.
In general, the proposed model produces superior and consistent QoE prediction performance in different situations with and without rebuffering events.
Patterns \#1-\#3 in Fig. \ref{fig:ResultLfovia}, \ref{fig:ResultLive}, and \ref{fig:ResultNetflix} show that the proposed model can effectively capture the effect of rebuffering events on the user's subjective QoE.
Especially, even the rebuffering event repeatedly occurs as illustrated in pattern \#3 in Fig. \ref{fig:ResultLfovia} and patterns \#2, \#3 in Fig. \ref{fig:ResultLive}, the QoE predictions still correlate well with the subjective QoE.
Meanwhile, pattern \#0 in Fig. \ref{fig:ResultLfovia} and pattern \#1 in Fig. \ref{fig:ResultNetflix} show some fluctuations in the predicted QoE.
However, the amplitudes of these fluctuations are small and the varying trends in the subjective QoE are still adequately captured by the proposed model.

The QoE prediction performance results over each database in comparison with existing models are shown in the Tables \ref{tbl:Accuracy_LFOVIA}, \ref{tbl:Accuracy_LiveMobileStall} and \ref{tbl:Accuracy_LiveNetflix}.
From these tables, it is revealed that the \Model{} outperforms the existing QoE models within all the criteria, especially in terms of RMSE.
Moreover, the accuracy produced by \Model{} is consistent across the databases, thus marking it as an efficient comprehensive model.
The results illustrate that the \Model{} architecture is capable of capturing the complex inter-dependencies and non-linear relationships among QoE influence factors.
Interestingly, there is a significant improvement in QoE prediction accuracy when comparing with \ModelTCN{}.
It means that the enhancements in the proposed architecture have made the model more suitable for QoE prediction.

On mobile devices (i.e., Android smartphones), the QoE prediction accuracy of the proposed \Model{} is assessed in comparison with \ModelTCN{} and LSTM-QoE \cite{QoEModel_LSTM}.
The results are shown in Table \ref{tbl:Accuracy_ComparePCvsMobile}.
Accordingly, when performing on different platforms (personal computers and mobile devices), the QoE prediction accuracy of both \Model{} and \ModelTCN{} remains unchange.
However, the LSTM-QoE, when performing on mobile devices, suffers from a significant loss in the QoE prediction accuracy.
This is because the precision of floating-point numbers is handled differently in different processors.


\subsection{Computational Complexity}
\label{sec:EvaluationComplexity}
  In this subsection, the computational complexity of the proposed model on personal computers and mobile devices is investigated.
  The purpose is to show the effectiveness of the \Model{} on both high and low computational devices in comparison with baseline methods including \ModelTCN{} and LSTM-QoE \cite{QoEModel_LSTM}.
  These models are trained and tested on the LFOVIA Video QoE Database with a training:test ratio of 80:20.

  \subsubsection{Evaluation Settings}
  \label{sec:EvaluationComplexity_Settings}
  For running these deep learning-based QoE prediction models, a personal computer running 18.04 Ubuntu LTS with an Intel i7-8750H @ 2.20GHz and 16GB RAM system is used.
On the Android side, Sony Xperia XA2, which runs Android 9.0 and possesses a Qualcomm Snapdragon 630 64-bit ARM-based octa-core system on a chip, is used.
Its CPU clock speed varies between 1.8-2.2 GHz depending on the core being used.
The internal memory of this smartphone is 3GB LPDDR4 RAM.
It should be noted that the GPU computation power is not utilized both on the personal computer and on Android smartphones.
  
  \subsubsection{Evaluation Criteria}
  \label{sec:EvaluationComplexity_Criteria}
  To conduct the evaluation on personal computers, the following four evaluation metrics are considered:
\begin{itemize}
    \item Inference time: the time taken to predict the user QoE $\widehat{y}_t$ at any given time instant $t$.
    \item Model size: the storage size of the trained model on the hard drive.
    \item FLOPs: the number of operations performed.
    \item Number of Parameters: number of learnable parameters in the model.
\end{itemize}

On the Android smartphones, after the conversion from Keras to TensorFlow model as described in Section \ref{sec:EvaluationAccuracy_Settings}, the model is for inference only since all the learnable parameters were converted to constants.
Therefore, the complexity of the \Model{} was compared with the others on Android smartphones using only two metrics: 1) Inference time and 2) Model size.
  
  \subsubsection{Results}
  \label{sec:EvaluationComplexity_Results}
  \begin{table*}[tb]
  \centering
    \begin{tabular}{|l|c|c|c|c|}
    \hline
    \multicolumn{1}{|c|}{} & \begin{tabular}[c]{@{}c@{}}Inference Time\\ (ms)\end{tabular} & \begin{tabular}[c]{@{}c@{}}Model Size\\ (kB)\end{tabular} & FLOPs   & \begin{tabular}[c]{@{}c@{}}Number of\\ Parameters\end{tabular} \\ \hline
    \Model      & \textbf{0.673} &  82.08 & 175,498 &  9,605 \\ \hline
    \ModelTCN   & 0.856 & 145.86 & 253,076 & 13,766 \\ \hline
    LSTM-QoE \cite{QoEModel_LSTM}    & 1.996 &  \textbf{41.18} &  \textbf{30,953} &  \textbf{6,364} \\ \hline
  \end{tabular}

  \caption{Computational complexity of the \Model{} on the personal computer.}
  \label{tbl:Complexity_PC}
\end{table*}

\begin{table*}[tb]
  \centering
  \begin{tabular}{|l|c|c|}
    \hline
    \multicolumn{1}{|c|}{} & \begin{tabular}[c]{@{}c@{}}Inference Time\\ (ms)\end{tabular} & \begin{tabular}[c]{@{}c@{}}Model Size\\ (kB)\end{tabular} \\ \hline
    \Model      & \textbf{0.581} & \textbf{56.222} \\ \hline
    \ModelTCN   & 0.640 & 88.67 \\ \hline
    LSTM-QoE \cite{QoEModel_LSTM}    & 1.550 & 77.04 \\ \hline
  \end{tabular}
  \caption{Computational complexity of the \Model{} on Android smartphones.}
  \label{tbl:Complexity_Mobile}
\end{table*}

Table \ref{tbl:Complexity_PC} and \ref{tbl:Complexity_Mobile} show the computational complexity results of the proposed \Model{} compared to the \ModelTCN{} and LSTM-QoE.
In general, the \Model{} requires a higher number of parameters and FLOPs in comparison with LSTM-QoE to achieve higher accuracy.
Although the FLOPs of the proposed model are larger, the inference time is 3 times faster than the LSTM-QoE model both on the personal computer and the Android smartphone.
This indicates that the proposed model can efficiently leverage the power of parallel computation to boost up the computing speed.
It can be seen from Table \ref{tbl:Complexity_PC} that the architecture complexity of \ModelTCN{} is extremely higher than our proposed \Model{} model in terms of number of parameters and FLOPs.
However, the accuracy of \ModelTCN{} is not quite comparable with the \Model{} as shown in Table \ref{tbl:Accuracy_ComparePCvsMobile}.
It proves that the proposed improvement adapted on the original TCN architecture allow \Model{} to effectively capture the complex temporal dependencies in a sequential QoE data.


\subsection{Overall Performance}
\label{sec:EvaluationOverall}
Accurate and efficient QoE prediction models provide important benefits to the deployment and operation of video streaming services on different viewing devices.
As shown in subsection \ref{sec:EvaluationAccuracy} and \ref{sec:EvaluationComplexity}, the proposed model \Model{} can achieve not only the state-of-the-art QoE prediction accuracy but also the reduction on computational complexity.
Therefore, the \Model{} can be an excellent choice for future QoE prediction systems or QoE-driven video streaming mobile applications.

\section{Discussion}
\label{sec:Discussion}
According to the above-mentioned evaluation results, it can be seen that the proposed model completely outperforms \ModelTCN{} where the original TCN architecture is adopted in the QoE prediction task.
Thereby, it generally demonstrates the efficiency of the proposed improvements upon the original TCN architecture in QoE prediction for video streaming services.
In the following subsections, the effects of the improvements are discussed in detail.


\subsection{Effects of comprising causal convolutions and dilated causal convolutions}

Different from TCN \cite{Network_TCN} architecture, the proposed architecture has an initial causal convolution instead of a dilated causal convolution, as shown in Fig. \ref{fig:ProposedArchitecture}.
Unlike dilated causal convolution, a causal convolution with denser filters is more effective in extracting the local dependencies among adjacent time-steps.
However, a stack of causal convolutions dramatically increases the model complexity.
Therefore, we combine causal convolutions with dilated causal convolutions to achieve desirable prediction accuracy, while eliminating the complexity possibly caused by only utilizing causal convolutions in the architecture.
As a result, the proposed model can effectively capture the temporal dependencies among adjacent and far-apart time-steps in the sequential QoE data, providing a better QoE prediction accuracy, especially in terms of RMSE.

Moreover, it can be seen from Tables \ref{tbl:Complexity_PC} and \ref{tbl:Complexity_Mobile} that the FLOPs of the proposed model are larger than those of LSTM-QoE.
The reason is that the convolution layers require more operations for performing convolution between a number of filters and the input time series.
However, the proposed model runs faster than the baseline models on both personal computers and Android smartphones.
This indicates that the convolution operations are fully parallelized, leading to real-time QoE prediction advantages.


\subsection{Effects of simplifying the residual block and using SeLU}

To simplify the residual block, we adopt only one dilated causal convolution in the residual block instead of two as in the original TCN architecture (as illustrated in Fig. \ref{fig:TCN_residualBlock} and Fig. \ref{fig:ProposedResidualBlock}).
The reason behind this is the fact that the sequential QoE data is much simpler than the preferred data of TCN \cite{Network_TCN} (i.e., speech signal data).
Therefore, two dilated causal convolution layers can make the model easily suffers from overfitting and reduces the QoE prediction accuracy.
Reducing the number of dilated causal convolutions in the residual block helps the proposed model to be easily trained and reduce overfitting.
Furthermore, SeLU \cite{SeLU} activation function also enables the model to learn faster and converge better to the optimal values, subsequently improving the QoE prediction accuracy.

In terms of computational complexity, observing from Tables \ref{tbl:Complexity_PC} and \ref{tbl:Complexity_Mobile}, it is obvious that these improvements in the residual block tremendously reduced the number of parameters compared to the one in the original TCN architecture \ModelTCN{}.
Thereby, the \Model{} can produce smaller model size and FLOPs, faster training and inference times.

In summary, the improvements in the proposed architecture help provide a more stable, accurate and light-weight QoE prediction model.

\section{Conclusion}
\label{sec:Conclusion}
In this paper, the CNN-QoE model is proposed for continuous QoE prediction.
The proposed model introduces multiple improvements to the original TCN model to leverage its strengths and eliminate its drawbacks in the QoE prediction task for video streaming services.
The comprehensive evaluation across different QoE databases on both personal computers and mobile devices demonstrates that CNN-QoE produces superior performance in terms of QoE prediction accuracy and computational complexity.
Accordingly, CNN-QoE outperforms the state-of-the-art models.
These results validate the robustness of the proposed model in real-time QoE prediction on different platforms and devices.


\bibliography{references.bib}
\bibliographystyle{IEEEtran}


\end{document}